\documentclass[12pt]{article}
\usepackage{latexsym}
\usepackage[cp1250]{inputenc}

\title{Heisenberg algebra for restricted Landau problem}
\author{Krzysztof Andrzejewski\thanks{supported by the {\L}\'od\'z University grant No. 690.},\\ 
 Pawe\l\  Ma\'slanka$^*$ \\
Department of Theoretical Physics II \\
University of {\L}\'od\'z \\
Pomorska 149/153, 90 - 236 {\L}\'od\'z/Poland.}
\date{}

\begin{document}
\maketitle
\begin{abstract}
Algebraic derivation of modified Heisenberg commutation rules for restricted Landau problem is given.
\end{abstract}

\newpage
It appears that the noncommutative geometry should play an important role in our attempts to understand
the structure of space-time at short distances/high energies. Therefore, it seems 
reasonable to study how the fundamental notions of noncommutative geometry arise in simple physical settings.
 In particular, much attention has been paid to problem of motion of charged particles in two dimensions
under the influence of constant perpendicularly applied magnetic field, \cite{b1} - \cite{b7} (see also \cite{b8}). 
The noncommutativity of coordinates appears here naturally after restricting the system to the lowest Landau
level; for the most recent discussion see \cite{b13}. This can be seen either by skipping the kinetic term 
in the lagrangian and quantizing the resulting
constrained system by Dirac method or by saturating the commutation rule for coordinates by the
states belonging to the lowest Landau level only. The generalization to the higher Landau 
level has been also given \cite{b9}, \cite{b10}. It appears that, when higher levels are included the commutator
of coordinates is no longer a c-number.

 In the present brief note we give a general, purely algebraic derivation of commutation
rules for basic canonical variables for the theory restricted to arbitrary finite number of Landau levels. 
The only things we are
using are the Heisenberg commutation relations and the form of the energy spectrum.

Our starting point is the hamiltonian describing the planar motion of mass $m$\ charged particle in perpendicular 
magnetic field $B$\
\begin{eqnarray}
H=\frac{1}{2m}(p_i+\frac{eB}{2c}\varepsilon_{ik}x_k)^2;\label{w1}
\end{eqnarray}
here we have chosen the rotational invariant gauge, $A_i=-\frac{B}{2}\varepsilon_{ik}x_k$. 
The spectral decomposition of $H$\ reads
\begin{eqnarray}
H=\sum_{n=0}^{\infty}\hbar\omega (n+\frac{1}{2})P_n\label{w2}
\end{eqnarray}
where $\omega=\frac{eB}{mc}$\ and $P_n$\ is the projector on $n$-th level (which is infinite-dimensional subspace). 
$P_n$\ can be expressed in terms of $H$\ as follows:
\begin{eqnarray}
P_n=\frac{4}{ \pi (2n+1)}\frac{sin\left((n+\frac{1}{2})\pi (\frac{H}{(n+\frac{1}{2})\hbar \omega}-1)\right)}{(\frac{H}{(n+
\frac{1}{2})\hbar \omega}-1)(\frac{H}{(n+\frac{1}{2})\hbar \omega}+1)}\label{w3}
\end{eqnarray}
One can easily check that the r. h. s. of eq. (\ref{w3}) is well defined. In what follows we shall use some simple
relations listed below:
\begin{eqnarray}
&&[H,\;x_i]=-\frac{i\hbar}{m}(p_i+\frac{m\omega}{2}\varepsilon_{ik}x_k)\equiv -i\hbar\dot{x}_i \nonumber \\
&&[H,\;[H,\;x_j]]=-i\hbar\omega\varepsilon_{jk}[H,\;x_k] \nonumber \\
&&[H,\;p_i-\frac{m\omega}{2}\varepsilon_{ik}x_k]=0\label{w4}\\
&&[[H,\;x_k],[H,\;x_l]]=-\frac{i\hbar^3}{m}\omega\varepsilon_{kl} \nonumber \\
&&\sum_{k=1}^2[H,\;x_k][H,\;x_k]=-\frac{2\hbar^2}{m}H\nonumber
\end{eqnarray}
Note, that the motion described by the hamiltonian (\ref{w1}) is purely harmonic with frequency $\omega$. Therefore,
if $A$\ is any operator linear in $x_i$'s and $p_i$'s, $[H,\;A]$\ connects only the eigenstates of $H$\ differing by 
$\hbar\omega $,
i.e. the only nonvanishing operators are  $P_n[H,\;A]P_{n+1}$\ and $P_{n+1}[H,\;A]P_n$. The followig relation is
a direct consequence of the last eq. (\ref{w4})
\begin{eqnarray}
\sum_{k=1}^2P_n[H,\;x_k](P_{n+1}+P_{n-1})[H,\;x_k]P_n=-\frac{2\hbar^3\omega}{m}(n+\frac{1}{2})P_n\label{w5}
\end{eqnarray}
On the other hand note that 
\begin{eqnarray}
&&\sum_{k=1}^2P_n[H,\;x_k](\hbar\omega (n+\frac{3}{2})P_{n+1}+\hbar\omega(n-\frac{1}{2})P_{n-1})[H,\;x_k]P_n=\nonumber \\
&&=\sum_{k=1}^2P_n[H,\;x_k]H[H,\;x_k]P_n=\nonumber \\
&&=\sum_{k=1}^2P_n[H,\;x_k][H,\;[H,\;x_k]]P_n-\frac{2\hbar^4\omega^2}{m}(n+\frac{1}{2})^2P_n=\label{w6} \\
&&=-i\hbar\omega\sum_{k,\;l=1}^2\varepsilon_{kl}P_n[H,\;x_k][H,\;x_l]P_n-\frac{2\hbar^4\omega^2}{m}(n+\frac{1}{2})^2P_n=\nonumber \\
&&=-\frac{i\hbar\omega}{2}\sum_{k,\;l=1}^2\varepsilon_{kl}P_n[[H,\;x_k],[H,\;x_l]]P_n-\frac{2\hbar^4\omega^2}{m}(n+\frac{1}{2})^2P_n=\nonumber \\
&&=-\frac{2\hbar^4\omega^2}{m}\left(\frac{1}{2}+(n+\frac{1}{2})^2\right)P_n\nonumber
\end{eqnarray}
From eqs. (\ref{w5}) and (\ref{w6}) one finds
\begin{eqnarray}
&&\sum_{k=1}^2P_n[H,\;x_k]P_{n+1}[H,\;x_k]P_n=-\frac{\hbar^3\omega}{m}(n+1)P_n\label{w7} \\
&&\sum_{k=1}^2P_n[H,\;x_k]P_{n-1}[H,\;x_k]P_n=-\frac{\hbar^3\omega}{m}nP_n\nonumber
\end{eqnarray}
Finally, we shall need the following relations
\begin{eqnarray}
&&[f(H),\;B]P_n=\frac{f(H)-f((n+\frac{1}{2})\hbar\omega )}{H-(n+\frac{1}{2})\hbar\omega}[H,\;B]P_n\label{w8} \\
&&P_n[f(H),\;B]=P_n[H,\;B]\frac{f(H)-f((n+\frac{1}{2})\hbar\omega)}{H-(n+\frac{1}{2})\hbar\omega}\nonumber
\end{eqnarray}
which hold for any operator $B$\ and any analytic function $f$. They can be proven by expanding $f(H)$\ in power series and using 
$P_nH=HP_n=\hbar\omega (n+\frac{1}{2})P_n$.

Let us now consider the theory obtained from the hamiltonian (\ref{w1}) by imposing the cutoff $E\leq \hbar\omega (N+\frac{1}{2})$. 
Let $\Pi_N=\sum_{n=0}^NP_n$\ be the projection operator on the relevant subspace. For any operator $B$\ let
\begin{eqnarray}
\hat{B}=\Pi_nB\Pi_n\label{w9}
\end{eqnarray}
be its counterpart in cut-off theory. We are interested in commutation rules between cut-off operators; in particular, we would like to 
calculate the commutation rules between basic dynamical variables. Let us start with $[\hat{x}_i,\;\hat{x}_j]$. Using $[x_i,\;x_j]=0$\ and
$\Pi _N[x_i,\;\Pi_N]\Pi_N=0$\ we find
\begin{eqnarray}
[\hat{x}_i,\;\hat{x}_j]=\Pi_N[\Pi_N,\;x_i][\Pi_N,\;x_j]\Pi_N-(i\leftrightarrow j)\label{w10}
\end{eqnarray}
From eq. (\ref{w8}) one gets
\begin{eqnarray}
[\Pi_N,\;x_i]\Pi_N=\frac{1}{H-(N+\frac{1}{2})\hbar\omega}(\Pi_N-1)[H,\;x_i]P_N\equiv g(H)[H,\;x_i]P_N\label{w11}
\end{eqnarray}
so that eq. (\ref{w10}) can be rewritten as 
\begin{eqnarray} 
&&[\hat{x}_i,\;\hat{x}_j]=P_N[H,\;x_i]g^2(H)[H,\;x_j]P_N-(i\leftrightarrow j)= \label{w12} \\
&&=P_n\left[[H,\;x_i],[H,\;x_j]\right]g^2(H)P_N+\left(P_N[H,\;x_i][g^2(H),[H,\;x_j]]P_N-(i\leftrightarrow j)\right)\nonumber
\end{eqnarray}
In the first term on r. h. s. of eq. (\ref{w12}) we find
\begin{eqnarray}
g^2(H)P_N=g^2\left((N+\frac{1}{2})\hbar\omega\right)P_N=\frac{1}{(2N+1)^2\hbar^2\omega^2}P_N\label{w13}
\end{eqnarray}
while for the second term we use again eq. (\ref{w8}) to get
\begin{eqnarray}
&&P_N[H,\;x_i]\frac{g^2(H)-g^2((N+\frac{1}{2})\hbar\omega )}{H-(N+\frac{1}{2})\hbar\omega}[H,\;[H,\;x_j]]P_n-(i\leftrightarrow j)=\nonumber \\
&&=i\hbar\omega\varepsilon_{ij}\sum_{k=1}^2P_N[H,\;x_k]\frac{g^2(H)-g^2((N+\frac{1}{2})\hbar\omega )}{H-(N+\frac{1}{2})\hbar\omega}[H,\;x_k]P_n\label{w14}
\end{eqnarray}
Inserting $I=\sum_{n=0}^{\infty}P_n$\ one can write the last expression in the form
\begin{eqnarray}
&&i\hbar\omega\varepsilon_{ij}\sum_{k=1}^2P_N[H,\;x_k]\left(\frac{g^2((N+\frac{3}{2})\hbar\omega )-g^2((N+\frac{1}{2})\hbar\omega )}{\hbar \omega}P_{N+1}+\right.\nonumber \\
&&\left.+\frac{g^2((N-\frac{1}{2})\hbar\omega )-g^2((N+\frac{1}{2})\hbar\omega )}{-\hbar\omega}P_{N-1}\right)[H,\;x_k]P_N=\nonumber \\
&&=-\frac{i\hbar\varepsilon_{ij} }{m\omega (2N+1)^2}(4N(N+1)^2+N)P_N\label{w15}
\end{eqnarray}
where we have used (\ref{w7}) and (\ref{w11}).\\
Collecting all equations starting from eq. (\ref{w12}) we arrive at the following result
\begin{eqnarray}
[\hat{x}_i,\;\hat{x}_j]=-\frac{i\hbar}{m\omega}(N+1)\varepsilon_{ij}P_N\label{w16}
\end{eqnarray}
In a similar way one can compute the remaining basic commutation rules
\begin{eqnarray}
&&[\hat{p}_i,\;\hat{p}_j]=-\frac{i\hbar m\omega}{4}(N+1)\varepsilon_{ij}P_N \label{w17}\\
&&[\hat{x}_i,\;\hat{p}_j]=i\hbar(1-\frac{1}{2}(N+1))\delta_{ij}P_N\nonumber
\end{eqnarray}
Eqs. (\ref{w16}), (\ref{w17}) provide the basic algebra for cut-off theory. Keeping in mind that $P_N$\ is given by eq. (\ref{w3}) we see that the Heisenberg 
algebra is in this case a kind of $W$-algebra. The validity of the algebra (\ref{w16})-(\ref{w17}) can be checked by explicit construction of the space of
states \cite{b11}. We define the anihilation and creation operators $a_{\pm},\;a_{\pm}^+$:
\begin{eqnarray}
&&a_+=\frac{1}{\sqrt{2m\omega\hbar}}(p_1+ip_2)-\frac{i}{2}\sqrt{\frac{m\omega}{2\hbar}}(x_1+ix_2)\nonumber \\
&&a_-=\frac{1}{\sqrt{2m\omega\hbar}}(p_1-ip_2)-\frac{i}{2}\sqrt{\frac{m\omega}{2\hbar}}(x_1-ix_2)\nonumber \\
&&a_+^+=\frac{1}{\sqrt{2m\omega\hbar}}(p_1-ip_2)+\frac{i}{2}\sqrt{\frac{m\omega}{2\hbar}}(x_1-ix_2)\label{w18} \\
&&a_-^+=\frac{1}{\sqrt{2m\omega\hbar}}(p_1+ip_2)-\frac{i}{2}\sqrt{\frac{m\omega}{2\hbar}}(x_1+ix_2)\nonumber 
\end{eqnarray}
The hamiltonian (\ref{w1}) takes now the form
\begin{eqnarray}
H=\hbar\omega (a_+^+a_++\frac{1}{2})\label{w19}
\end{eqnarray}
while the angular momentum reads
\begin{eqnarray}
L=\hbar (a_-^+a_--a_+^+a_+)\label{w20}
\end{eqnarray}
We see that $a_-,\;a^+_-$\ produce the states of the same energy but different angular momentum.

Our cut-off space of states,(see, also, \cite{b12}), is spanned by the vectors $\mid n_+,\;n_->,\;n_+=0,\;1,\;...,\;N,\;n_-=0,\;1,\;2,\;...$. The modified
commutation rules read
\begin{eqnarray}
&&[a_-,\;a_-^+]=1\nonumber\\
&&[a_+,\;a_+^+]=1-(N+1)P_N\label{w21}\\
&&[a_+,\;a_-]=[a_+^+,\;a_-^+]=[a_+^+,\;a_-]=[a_+,\;a^+]=0\nonumber
\end{eqnarray}
and
\begin{eqnarray}
P_N\mid n_+,\;n_->=\delta_{N,\;n_+}\mid n_+,\;n_->\label{w22}
\end{eqnarray}
Using eqs. (\ref{w18}) and (\ref{w21}) one easily verifies the validity of commutation rules (\ref{w16}) 
and (\ref{w17}).
\\

{\bf Acknowledgment}
\\

We would like to thank prof. P. Kosi\'nski for interesting discussions.

\end{document}